\documentclass[11pt]{article}

\usepackage[margin=1in]{geometry}
\usepackage[utf8]{inputenc}
\usepackage[T1]{fontenc}
\usepackage[numbers,sort&compress]{natbib}
\usepackage[hidelinks]{hyperref}
\usepackage{url}
\usepackage{enumitem}
\usepackage[table]{xcolor}
\usepackage{tikz}
\usetikzlibrary{positioning, arrows.meta, shapes.geometric, calc}
\usepackage{listings}

\lstdefinestyle{yamlstyle}{
  basicstyle=\ttfamily\footnotesize,
  keywordstyle=\color{blue!70!black},
  stringstyle=\color{purple!70!black},
  commentstyle=\color{gray},
  numbers=none,
  frame=single,
  framesep=2pt,
  rulecolor=\color{gray!50},
  breaklines=true,
  showstringspaces=false,
  tabsize=2,
  morekeywords={schema_version, allowRetake, initialPageId, pages, components, props, buttons, action, onEnter, end, agents, matchmaking, type, id, target, conditions, stateKey, assignmentType, poolId, num_users, timeoutSeconds, timeoutTarget, model, system, tools, layout, mode, editableBy, initialContent, placeholder, parameters, properties, required, description, name, src, read, write, height},
  keepspaces=true,
}

\usepackage{booktabs}
\usepackage{array}
\usepackage{amssymb}
\usepackage{longtable}
\title{Pairit: A Platform for Live Experiments on \\ Human--AI Collaboration}

\author{%
  Harang Ju\thanks{Carey Business School, Johns Hopkins University. \href{mailto:harang@jhu.edu}{harang@jhu.edu}.}
  \and
  Sinan Aral\thanks{Sloan School of Management, Massachusetts Institute of Technology. Correspondence: \href{mailto:sinan@mit.edu}{sinan@mit.edu}.}
}

\date{}

\begin{document}
\maketitle

\begin{abstract}
  Organizational design in the era of artificial intelligence requires experimental methods that can test how human--AI groups coordinate, delegate, and make decisions. Programmable platforms coordinate live human-to-human sessions or real-time human--AI chat, but researchers cannot easily declare experiment protocols in which AI participants both communicate and act on shared work within one auditable configuration. Here we introduce Pairit, an online platform that facilitates the design, testing, and deployment of experiments that test human--AI organizational designs and interventions. Through a single YAML configuration file, researchers declare an executable experiment graph---pages, routing, randomization, matchmaking, chat, shared workspaces, server-hosted agents, surveys, timers, and custom HTML components---and combine any number of humans and AI agents in live sessions. We have validated the feasibility of the platform through multiple live deployments, including peer-reviewed published studies, capturing high-resolution process traces of communication, negotiation, and collaborative work in live human--AI dyads. By representing complex interactive protocols as standardized, auditable configuration files, Pairit provides reusable infrastructure for specifying, deploying, and sharing live human--AI organizational experiments.
\end{abstract}

\section{Introduction}

Researchers run experiments to causally test design choices in human--AI teams. Programmable platforms such as oTree \citep{chen2016otree} and Empirica \citep{almaatouq2021empirica} coordinate live human-to-human sessions, and newer systems such as Deliberate Lab \citep{qian2025deliberatelab} support real-time human--AI group chat. Yet researchers still cannot easily declare experiment protocols in which AI agents converse, co-edit shared documents, and take protocol-defined actions within one auditable configuration that also specifies team assignment, communication channels, and routing. Without that capability, researchers cannot easily and systematically vary team composition, live communication, and how agents collaborate during tasks. Because each new design requires custom software, live experiment protocols rarely become shareable research objects other scholars can inspect or run.

\begin{figure}[ht]
\centering
\begin{tikzpicture}[
  human/.style={circle, draw, thick, minimum size=0.62cm, inner sep=0pt, font=\tiny\bfseries},
  agent/.style={rectangle, draw, thick, rounded corners, minimum width=0.82cm, minimum height=0.5cm, inner sep=1pt, font=\tiny\bfseries, fill=black!8},
  link/.style={thick, gray!45!black},
  card/.style={draw=gray!35, rounded corners=3pt, fill=gray!4, minimum width=4.48cm, minimum height=1.62cm, anchor=north west},
  cardtitle/.style={font=\tiny\bfseries, text=black!75, anchor=north west},
  chip/.style={draw=gray!50, fill=white, rounded corners=2pt, font=\tiny, inner sep=1.6pt, minimum height=0.36cm, minimum width=1.92cm, align=center},
]
  \node[human] (a1) at (0.0,0) {H};
  \node[human] (a2) at (1.25,0) {H};
  \draw[link] (a1) -- (a2);
  \node[font=\scriptsize\bfseries] at (0.625,0.85) {Human--human};
  \node[font=\tiny] at (0.625,-0.7) {Same live task};

  \node[human] (b1) at (3.85,0) {H};
  \node[agent] (b2) at (5.15,0) {AI};
  \draw[link] (b1) -- (b2);
  \node[font=\scriptsize\bfseries] at (4.5,0.85) {Human--AI};
  \node[font=\tiny] at (4.5,-0.7) {Randomized partner};

  \node[human] (c1) at (7.55,0.28) {H};
  \node[human] (c2) at (8.8,0.28) {H};
  \node[agent] (c3) at (8.175,-0.62) {AI};
  \draw[link] (c1) -- (c2);
  \draw[link] (c1) -- (c3);
  \draw[link] (c2) -- (c3);
  \node[font=\scriptsize\bfseries] at (8.175,0.85) {Group + AI};
  \node[font=\tiny] at (8.175,-1.15) {Facilitator in the group};

  \node[human] (n1) at (11.15,0.28) {H};
  \node[human] (n2) at (12.4,0.28) {H};
  \node[agent] (n3) at (13.35,-0.55) {AI};
  \draw[link] (n1) -- (n2);
  \draw[link] (n2) -- (n3);
  \node[font=\scriptsize\bfseries] at (12.15,0.85) {Side assist};
  \node[font=\tiny] at (12.15,-1.15) {Rewrite in a bargain};

  \node[human] (d1) at (0.0,-2.85) {H};
  \node[human] (d2) at (1.25,-2.85) {H};
  \node[human] (d3) at (0.625,-3.75) {H};
  \draw[link] (d1) -- (d2);
  \draw[link] (d1) -- (d3);
  \draw[link] (d2) -- (d3);
  \node[font=\scriptsize\bfseries] at (0.625,-2.15) {Small group};
  \node[font=\tiny] at (0.625,-4.4) {Any group size};

  \node[agent] (e1) at (3.85,-3.15) {AI};
  \node[agent] (e2) at (5.15,-3.15) {AI};
  \draw[link] (e1) -- (e2);
  \node[font=\scriptsize\bfseries] at (4.5,-2.15) {AI-only};
  \node[font=\tiny] at (4.5,-4.4) {Baseline arm};

  \node[human] (f1) at (7.55,-3.15) {H};
  \node[agent] (f2) at (8.8,-3.15) {AI};
  \draw[link] (f1) -- (f2);
  \node[font=\tiny, gray!50!black] at (8.175,-3.75) {from intake};
  \node[font=\scriptsize\bfseries] at (8.175,-2.15) {Pairing};
  \node[font=\tiny] at (8.175,-4.4) {Persona from scores};

  \node[human] (g1) at (11.15,-2.7) {H};
  \node[human] (g2) at (12.4,-2.7) {H};
  \node[human] (g3) at (11.15,-3.7) {H};
  \node[agent] (g4) at (12.4,-3.7) {AI};
  \draw[link] (g1) -- (g2);
  \draw[link] (g1) -- (g3);
  \draw[link] (g2) -- (g4);
  \draw[link] (g3) -- (g4);
  \node[font=\scriptsize\bfseries] at (11.775,-2.15) {Mixed group};
  \node[font=\tiny] at (11.775,-4.4) {Any $n$ humans, $m$ agents};

  \node[font=\scriptsize\bfseries, anchor=west] at (-0.25,-5.55) {Built-in components};
  \node[card] (box_live) at (-0.25,-5.85) {};
  \node[cardtitle] at ($(box_live.north west)+(0.16,-0.10)$) {Live interaction \& AI};
  \node[chip] at ($(box_live.center)+(-1.04,-0.06)$) {Matchmaking};
  \node[chip] at ($(box_live.center)+(1.04,-0.06)$) {Chat};
  \node[chip] at ($(box_live.center)+(-1.04,-0.52)$) {Workspace};
  \node[chip] at ($(box_live.center)+(1.04,-0.52)$) {Agents};

  \node[card] (box_input) at (4.47,-5.85) {};
  \node[cardtitle] at ($(box_input.north west)+(0.16,-0.10)$) {Inputs \& custom embeds};
  \node[chip] at ($(box_input.center)+(-1.04,-0.06)$) {HTML};
  \node[chip] at ($(box_input.center)+(1.04,-0.06)$) {Survey};
  \node[chip] at ($(box_input.center)+(-1.04,-0.52)$) {Form};
  \node[chip] at ($(box_input.center)+(1.04,-0.52)$) {Media};

  \node[card] (box_flow) at (9.19,-5.85) {};
  \node[cardtitle] at ($(box_flow.north west)+(0.16,-0.10)$) {Experiment flow};
  \node[chip] at ($(box_flow.center)+(-1.04,-0.06)$) {Randomize};
  \node[chip] at ($(box_flow.center)+(1.04,-0.06)$) {Timer};
  \node[chip] at ($(box_flow.center)+(-1.04,-0.52)$) {Text};
  \node[chip] at ($(box_flow.center)+(1.04,-0.52)$) {Buttons};
\end{tikzpicture}
\caption{Example team compositions and built-in components. Any $n$ humans and $m$ AI agents can co-exist in one study.}
\label{fig:compositions}
\end{figure}
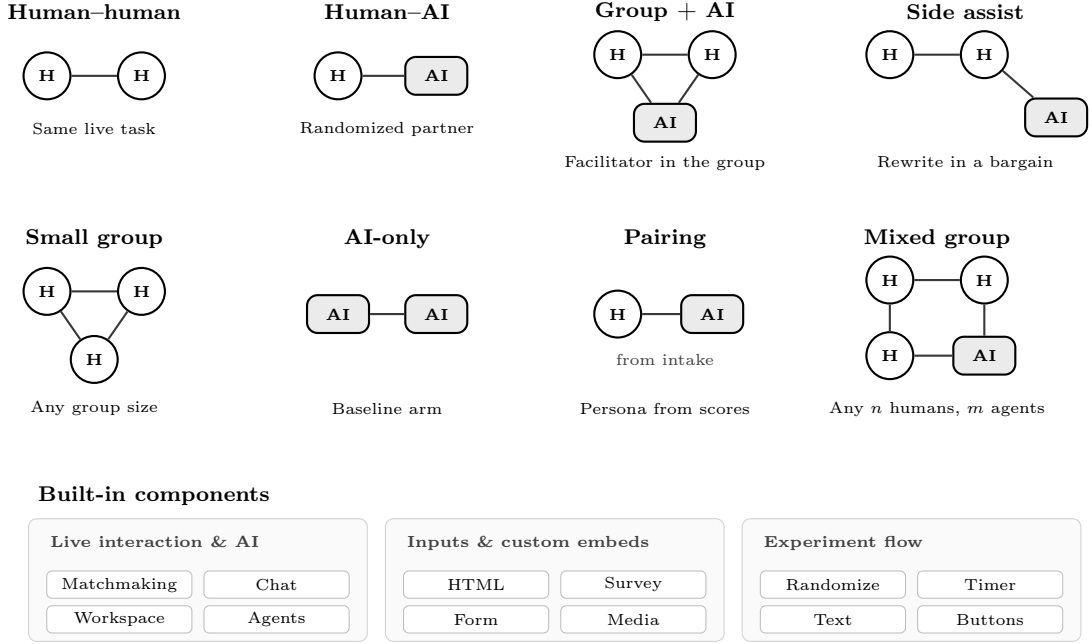

Here we introduce Pairit, which lets researchers declare live experiment protocols with any mix of human and AI participants in a single auditable configuration, run them as live sessions, and share them with other labs. Researchers can combine any number of humans and AI participants in one session (Figure~\ref{fig:compositions}), from human--human baselines to human--AI dyads, groups with embedded facilitators, and side-by-side negotiation assistants. Sessions combine matchmaking, chat, and shared workspaces, and server-hosted AI can join the conversation, edit shared documents, and execute protocol actions such as updating session state or ending a discussion. Researchers configure these roles dynamically---for example, setting persona prompts based on intake survey answers or activating assistance only during specific task stages---so organizational scholars can test how AI teammates alter communication, coordination, delegation, and collective performance in live environments.

\section{System Architecture and Declarative Experiment Graphs}

\begin{figure}[!ht]
\centering
\begin{minipage}[t]{0.31\textwidth}
\centering
{\small\textbf{A. Declarative config}}\par\vspace{4pt}
\begin{lstlisting}[style=yamlstyle, basicstyle=\ttfamily\tiny, numbers=none, frame=single, framesep=2pt, aboveskip=0pt, belowskip=0pt, xleftmargin=0pt, xrightmargin=0pt, breaklines=false]
pages:
  - id: match
    components:
      - type: matchmaking
  - id: discuss
    layout: split
    components:
      - type: chat
        props:
          agents: [facilitator]
      - type: live-workspace
matchmaking:
  - id: hire_pool
    num_users: 2
agents:
  - id: facilitator
    model: gpt-5-nano
\end{lstlisting}
\end{minipage}\hfill
\begin{minipage}[t]{0.05\textwidth}
\vspace{1.9cm}
\centering
{\Large$\rightarrow$}\\[-1pt]
{\tiny\itshape compile}
\end{minipage}\hfill
\begin{minipage}[t]{0.24\textwidth}
\centering
{\small\textbf{B. Compiled graph}}\par\vspace{0.85cm}
\begin{tikzpicture}[
  node distance=0.38cm,
  page/.style={rectangle, draw, rounded corners, minimum width=1.55cm, minimum height=0.5cm, font=\scriptsize, align=center, fill=white, inner sep=2pt},
  hidden/.style={page, fill=gray!18, dashed},
  endpage/.style={page, fill=black!8},
  flow/.style={->, >=stealth, thick, gray!50!black},
]
\node[page] (g_intro) {Intro};
\node[hidden, below=of g_intro] (g_match) {Match};
\node[page, below=of g_match] (g_chat) {Discuss};
\node[endpage, below=of g_chat] (g_end) {End};
\node[endpage, right=0.22cm of g_match] (g_fallback) {Timeout};

\draw[flow] (g_intro) -- (g_match);
\draw[flow] (g_match) -- (g_chat);
\draw[flow] (g_chat) -- (g_end);
\draw[flow] (g_match) -- (g_fallback);
\end{tikzpicture}
\end{minipage}\hfill
\begin{minipage}[t]{0.05\textwidth}
\vspace{1.9cm}
\centering
{\Large$\rightarrow$}\\[-1pt]
{\tiny\itshape render}
\end{minipage}\hfill
\begin{minipage}[t]{0.31\textwidth}
\centering
{\small\textbf{C. Participant experience}}\par\vspace{4pt}
\begin{tikzpicture}[font=\tiny]
\node[draw=gray!50, rounded corners=2pt, fill=blue!3, minimum width=2.22cm, minimum height=4.45cm, anchor=north west, inner sep=0pt] (chat) at (0,0) {};
\node[anchor=north west, font=\tiny\bfseries] at ($(chat.north west)+(0.16,-0.16)$) {Chat};
\node[draw=gray!40, rounded corners=1.5pt, fill=white, anchor=north west, align=left, inner sep=2pt, text width=1.78cm] at ($(chat.north west)+(0.16,-0.68)$) {\textbf{You:} I think Alice is the stronger fit.};
\node[draw=gray!40, rounded corners=1.5pt, fill=white, anchor=north west, align=left, inner sep=2pt, text width=1.78cm] at ($(chat.north west)+(0.16,-1.68)$) {\textbf{Partner:} Bob has more years on the team.};
\node[draw=gray!40, rounded corners=1.5pt, fill=teal!10, anchor=north west, align=left, inner sep=2pt, text width=1.78cm] at ($(chat.north west)+(0.16,-2.68)$) {\textbf{Facilitator:} What would tip the decision either way?};
\node[draw=gray!40, rounded corners=1.5pt, fill=white, anchor=south, minimum width=1.90cm, minimum height=0.32cm, inner sep=0pt] at ($(chat.south)+(0,0.16)$) {\fontsize{5}{6}\selectfont\hspace{3pt}Type a message\dots};

\node[draw=gray!50, rounded corners=2pt, fill=yellow!4, minimum width=2.22cm, minimum height=4.45cm, anchor=north west, inner sep=0pt] (ws) at ($(chat.north east)+(0.24,0)$) {};
\node[anchor=north west, font=\tiny\bfseries] at ($(ws.north west)+(0.16,-0.16)$) {Workspace};
\node[draw=gray!40, rounded corners=1.5pt, fill=white, anchor=north west, minimum width=1.90cm, minimum height=3.40cm, inner sep=0pt] (wstext) at ($(ws.north west)+(0.16,-0.68)$) {};
\node[anchor=north west, align=left, inner sep=0pt, text width=1.70cm] at ($(wstext.north west)+(0.10,-0.12)$) {%
Hire recommendation\\[4pt]
Alice: stronger design.\\
Bob: more tenure.\\[4pt]
We lean Alice because|};
\end{tikzpicture}
\end{minipage}
\caption{From a declared configuration to the participant experience. \textbf{(A)} Pages, matchmaking, and an AI facilitator live in one YAML file. \textbf{(B)} The file compiles to a graph of nodes and edges. \textbf{(C)} Participants see the rendered study with a facilitator and shared workspace.}
\label{fig:overview}
\end{figure}
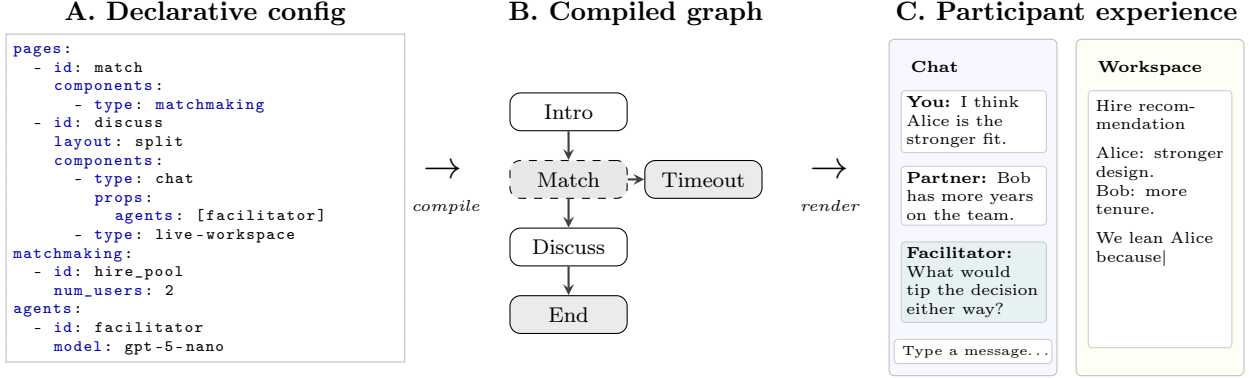

Pairit represents each experiment as a directed graph: pages are nodes, and participant actions and routing rules are edges (Figure~\ref{fig:overview}). Researchers declare this graph in a single YAML configuration file, specifying page flow, conditional routing, randomization, matchmaking, chat and workspace permissions, and agent models, prompts, and triggers. Under this declarative model, session variables track experimental state to drive conditional routing and real-time interface updates. Researchers can use built-in components for core primitives, including surveys, stimulus displays, timers, matchmaking pools, interactive chat rooms, and collaborative workspaces. During matchmaking, the platform can randomize team composition (human--human versus human--AI) and agent assignment within a condition. It then routes participants through private chats, multi-party discussions, or joint drafting environments, in which other participants or server-hosted AI agents chat or edit shared documents, and ends the collaboration when protocol conditions are met. Researchers configure these agents in the same file, selecting the model and writing system prompts, including conditional prompt blocks based on session state, and trigger rules that govern when agents speak or act. Collaborative workspaces support split-panel layouts with explicit permissions assigned to humans and agents. To support custom, experiment-specific interfaces, Pairit allows researchers to create and embed custom HTML components that integrate with the data, flow, and logic of the platform. During live sessions, Pairit captures process traces, including timestamped chat messages, keystroke-level workspace revisions, and delegation choices, and exports them in CSV, JSON, and JSONL formats.

The declarative experiment graph decouples experimental design from software engineering. Established behavioral platforms, including oTree \citep{chen2016otree}, Empirica \citep{almaatouq2021empirica}, jsPsych \citep{deleeuw2015jspsych,deleeuw2023jspsychjoss}, and nodeGame \citep{balietti2017nodegame}, distribute study logic across multiple imperative code files. Survey tools such as Qualtrics eliminate custom programming, but cannot coordinate synchronous multi-party matchmaking, collaborative drafting, or AI participants that both converse and act on shared work within a declared protocol. Moreover, as detailed in the technical appendix, existing frameworks were not built to treat AI as first-class organizational actors that both converse and take protocol-defined action within a declared graph. Pairit specifies page flow, routing, randomization, matchmaking, chat, workspaces, and agent configuration in a single structured configuration. Peer reviewers can inspect the entire organizational architecture directly in the configuration file, while independent laboratories can inspect, adapt, and run published protocols by adjusting explicit parameters like group size and agent roles.

\section{Validation and Reproducibility}

Pairit has been validated through multiple live deployments in peer-reviewed research and working papers, including published work on human--AI collaboration \citep{ju2026personality}. Across these deployments, we randomized team composition between human--human and human--AI groups among 2,234 Prolific participants collaborating in synchronized pairs to produce advertising campaigns later tested in a live market \citep{ju2025collaborating}, then randomized AI personality through trait prompts within human--AI teams \citep{ju2026personality}. Process logs from these studies supported analyses of gender gaps, diversity collapse, and the jagged frontier \citep{ju2026gender,ju2026diversity,ju2026jagged}. We also embedded real-time AI assistance in dyadic negotiations among 2,500 participants to study the effects of different types of AI assistance \citep{white2026negotiation}. We thus demonstrate that our declarative architecture reliably coordinates complex, real-time interactions across diverse organizational settings.

Pairit is operational today, with comprehensive documentation, an interactive template library, and browser demos at \href{https://docs.pairium.ai/pairit/}{docs.pairium.ai/pairit}. Before deployment, the Manager CLI lints and compiles each configuration against a strict schema. Researchers also declare consent pages, disclosure rules, and logging choices in the same configuration. Researchers can run these templates immediately to preview surveys, randomized team assignments, and dyadic human--AI collaboration, or use AI coding agents with the public documentation to draft and revise YAML configurations. The source is public for inspection. The reproducible objects are public configurations and export schemas; other laboratories can inspect, adapt, and run protocol structure on the hosted platform without access to prior-study participant-level data. We designed this infrastructure to make multi-agent designs fully auditable and directly reproducible.

\section{Implications for Organizational Research}

Pairit enables new research on coordination, communication, and delegation in live human--AI teams. Until now, studying these dynamics meant building live infrastructure from scratch: synchronizing participants, routing conversation, managing shared work, and embedding agents that both speak and act. That work remains slow and design-intensive even with AI-assisted coding, which is why many live human--AI designs were rarely tested in experiments. Pairit turns that build work into a declarative experiment graph, so researchers can systematically vary who participates, what role AI plays, and when it intervenes. Researchers can now test facilitation, delegation, and bargaining assistance in repeatable, customizable, and shareable live experiments rather than one-off custom builds.

Pairit also makes experiment protocols auditable, shareable, and reproducible. Standard prose methods sections cannot fully specify live experiment protocols. Instead, researchers can publish executable configurations alongside papers, allowing peer reviewers and other laboratories to audit prompts, routing, timeouts, and agent settings directly. Other laboratories can then reproduce or adapt the protocol via shared lab links. Granular process logs supplement these configurations rather than replacing them, providing a complete record of the interaction history.

Moreover, declarative experiment graphs open a path to in silico pretesting. Researchers can pilot designs before live recruitment by populating declared protocols with simulated participants using language models \citep{horton2023silicus}. Because these simulations run on the identical configuration file used for live sessions, pretests remain directly comparable to field deployments. We are developing this extension in parallel.

\section{Outlook}

We are currently rolling out beta access to outside laboratories so other groups can run published human--AI protocols from shared configurations rather than building bespoke, live infrastructure. One open question is whether research groups would adopt executable configurations as routinely as survey instruments. Pairit makes human-AI experiments easily configurable and portable, but wider use still depends on review, recruitment, model access, and hosting choices beyond configuration of the Pairit protocol. The technical appendix discusses these execution risks in detail. The authors are co-founders of Pairium, which develops and maintains the Pairit platform.

\clearpage
\bibliographystyle{plainnat}
\bibliography{refs}

\clearpage
\appendix
\section*{\LARGE Appendix}
\vspace{0.5em}
\setlength{\parskip}{0.45em}
\setlength{\parindent}{1.2em}
\section{Specification and Runtime Architecture}

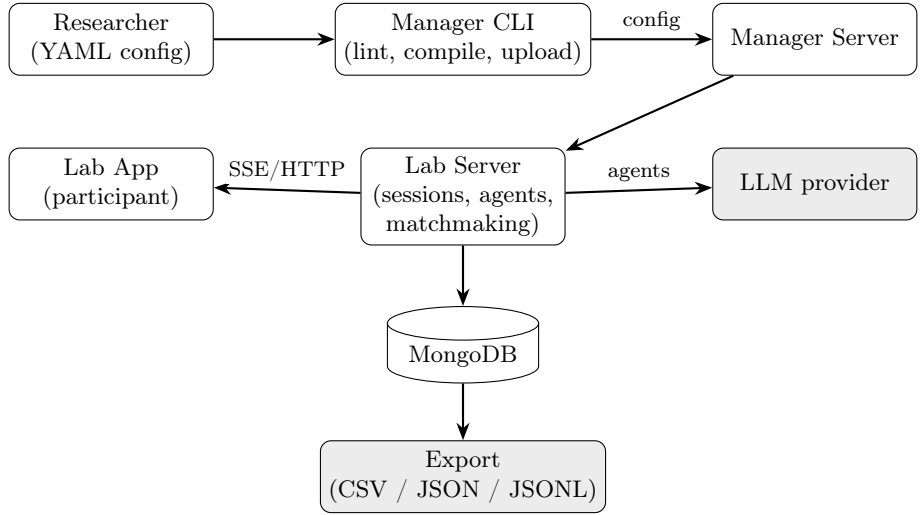
\begin{figure}[!ht]
\centering
\begin{tikzpicture}[
  node distance=0.85cm and 1.6cm,
  box/.style={rectangle, draw, rounded corners, minimum width=2.7cm, minimum height=0.95cm, inner sep=3pt, font=\footnotesize, align=center},
  data/.style={cylinder, draw, shape border rotate=90, aspect=0.28, minimum height=0.85cm, minimum width=2.0cm, font=\footnotesize, align=center, inner sep=2pt},
  ext/.style={box, fill=gray!15},
  edge/.style={->, >=Stealth, thick}
]
  \node[box] (author) {Researcher\\(YAML config)};
  \node[box, right=of author] (cli) {Manager CLI\\(lint, compile, upload)};
  \node[box, right=of cli] (manager) {Manager Server};

  \node[box, below=0.95cm of author] (labapp) {Lab App\\(participant)};
  \node[box, below=0.95cm of cli] (lab) {Lab Server\\(sessions, agents,\\matchmaking)};
  \node[ext, below=0.95cm of manager] (llm) {LLM provider};

  \node[data, below=0.8cm of lab] (db) {MongoDB};
  \node[ext, below=0.75cm of db] (export) {Export\\(CSV / JSON / JSONL)};

  \draw[edge] (author) -- (cli);
  \draw[edge] (cli) -- (manager) node[midway, above=2pt, fill=white, inner sep=1pt, font=\scriptsize] {config};
  \draw[edge] (manager) -- (lab);
  \draw[edge] (lab) -- (labapp) node[midway, above=2pt, fill=white, inner sep=1pt, font=\scriptsize] {SSE/HTTP};
  \draw[edge] (lab) -- (llm) node[midway, above=2pt, fill=white, inner sep=1pt, font=\scriptsize] {agents};
  \draw[edge] (lab) -- (db);
  \draw[edge] (db) -- (export);
\end{tikzpicture}
\caption{System architecture. Researchers author experiments as YAML configurations that the Manager CLI validates, compiles, and uploads. At runtime, the Lab Server reads the compiled configuration, coordinates sessions, matchmaking, and AI-agent calls, persists state and events to MongoDB, and streams updates to participant browsers via server-sent events. Exported data are produced from the underlying event and session stores.}
\label{fig:arch}
\end{figure}

Pairit structures an interactive experiment as a directed graph declared in a single YAML configuration file. Researchers author the study logic in this configuration, defining pages as graph nodes and participant actions as transition edges. The Manager CLI validates the configuration against a strict JSON schema, compiles the page flow into an execution graph, and uploads the compiled bundle to the Manager Server (Figure~\ref{fig:arch}). At runtime, the Lab Server ingests the compiled study graph to manage live participant sessions, synchronous matchmaking queues, and server-hosted AI agents. Participants interact with the study through the Lab App, a browser client that receives real-time state updates from the Lab Server over server-sent events and sends responses over standard HTTP requests. The platform persists all session state transitions, component events, chat turns, and document revisions to a MongoDB database for subsequent export into flat analysis files.

Graph navigation relies on strongly typed session state. When a participant selects an option or submits form input, transition rules evaluate the session state to determine the next page node. Pages also execute automated actions on entry through \texttt{onEnter} hooks. These hooks run background operations before rendering page content, such as assigning stratified conditions, computing intake scores, and routing participants into real-time matchmaking pools. When matchmaking succeeds, the server binds paired participants to a shared group session and advances them simultaneously to collaborative nodes. Researchers can inspect, test, and run these configurations immediately through the public documentation site at \href{https://docs.pairium.ai/pairit/}{docs.pairium.ai/pairit}, with source code hosted at \href{https://github.com/pairium/pairit}{github.com/pairium/pairit}.

\section{Component-First Architecture}

Pairit implements a component-first architecture where pages compose individual components to construct interactive experimental tasks. The framework treats matchmaking pools, chat rooms, collaborative workspaces, and server-managed agents as first-class components. The declarative YAML configuration mirrors the hierarchical structure of React JSX. Each component within a page can declare a \texttt{when} expression to govern conditional rendering. By evaluating these expressions against active session state, the server dynamically shows or hides components to guide participants through personalized experimental pathways.

\section{Built-in Component Reference}

Pairit provides a rich set of built-in components to construct diverse experimental designs. Table~\ref{tab:components} presents a summary of these components, their primary purposes, and their key capabilities. Every component supports conditional rendering via \texttt{when} expressions on \texttt{session\_state}.

\begin{longtable}{@{}>{\raggedright\arraybackslash}p{2.9cm} >{\raggedright\arraybackslash}p{3.2cm} >{\raggedright\arraybackslash}p{8.3cm}@{}}
\caption{Summary of built-in components and capabilities.} \label{tab:components} \\
\toprule
Component & Purpose & Key Capabilities \\
\midrule
\endfirsthead
\toprule
Component & Purpose & Key Capabilities \\
\midrule
\endhead
\midrule
\multicolumn{3}{r}{{Continued on next page}} \\
\endfoot
\bottomrule
\endlastfoot
\texttt{text} & Renders formatted text blocks inside participant pages. & Parses Markdown, compiles Mermaid.js diagrams, and interpolates dynamic \texttt{session\_state} variables via double-curly braces. \\
\midrule
\texttt{buttons} & Handles participant page transitions and action edges. & Supports \texttt{go\_to}, \texttt{next}, and \texttt{end} actions; conditional \texttt{branches} with default fallback; optional \texttt{onClick} event logging. \\
\midrule
\texttt{survey} & Gathers multi-question feedback and psychometric data. & Supports numeric, Likert (5/7), multiple-choice, multi-select, and free-text answers; paging; per-question \texttt{next} branching; optional media prompts; writes answers to \texttt{session\_state} on submit. \\
\midrule
\texttt{form} & Inputs custom structured data within general tasks. & Configures text, number, selector, and textarea fields; emits \texttt{onSubmit} and \texttt{onFieldChange} events for custom analytics. \\
\midrule
\texttt{chat} & Operates real-time conversational channels. & Runs solo or group rooms; attaches server-managed agents; shares or isolates message history across pages via \texttt{groupId}; streams agent replies in real time. \\
\midrule
\texttt{matchmaking} & Coordinates synchronous matchmaking queues. & Manages FIFO pools by \texttt{poolId} and \texttt{num\_users}; navigates to \texttt{onMatchTarget} or \texttt{timeoutTarget}; writes \texttt{chat\_group\_id} to session state; optional balanced or block assignment. \\
\midrule
\texttt{randomization} & Allocates experimental treatments on entry. & Performs random, balanced-random, or block assignment; runs invisibly via \texttt{onEnter} hooks or as a visible component; supports \texttt{scope: session} or \texttt{scope: group}; multiple \texttt{stateKey}s per page. \\
\midrule
\texttt{agents} & Runs autonomous conversational partners. & Connects to OpenAI or Anthropic via encrypted per-study keys; triggers on \texttt{every\_message}, \texttt{on\_join}, or \texttt{\{every: N\}}; \texttt{replyCondition} filters; conditional prompt blocks; guardrails and avatars; tools \texttt{assign\_state}, \texttt{end\_chat}, \texttt{write\_workspace}. \\
\midrule
\texttt{media} & Delivers sensory stimuli and instruction assets. & Plays images, video, and audio; captures \texttt{onPlay}, \texttt{onPause}, \texttt{onSeek}, \texttt{onComplete}, and \texttt{onError} events; assets uploaded via \texttt{pairit media upload}. \\
\midrule
\texttt{timer} & Enforces temporal constraints on experimental tasks. & Visible or invisible countdowns; warning thresholds; writes state and navigates via \texttt{action.branches} or \texttt{setState} on expiry. \\
\midrule
\texttt{live-workspace} & Enables collaborative document editing. & Freeform markdown or structured fields; \texttt{participant} or \texttt{group} scope; \texttt{editableBy} participant, agent, or both; pairs with chat via \texttt{layout: split}; agents read and edit via \texttt{write\_workspace}. \\
\midrule
\texttt{html} & Integrates fully bespoke client-side applications. & Sandboxed iframe with no network access; \texttt{read}/\texttt{write} state keys; \texttt{pairit.ready/setState/event/done} API; bundled upload with config; \texttt{required: true} blocks Next until task completion. \\
\end{longtable}

\paragraph{HTML Custom Interfaces.}
The \texttt{html} component stands as the central extension mechanism for bespoke experimental designs. Researchers can construct any custom user interface---such as real-time interactive games, slider scales, or specialized coordination tasks---as a single, self-contained HTML file. This file can incorporate pre-built React, Vue, or Svelte bundles. When researchers publish their study, the Manager CLI packages and uploads the HTML file alongside the core configuration using \texttt{pairit config upload}. The CLI linter automatically validates that the file exists locally, remains within a 1MB limit, and contains no syntactic errors.

These custom interfaces do not operate as isolated widgets. Instead, they integrate deeply with the platform services using the client-side \texttt{pairit} helper API. Through this API, the embedded page executes operations like \texttt{pairit.ready()} to signal initialization, \texttt{pairit.setState(keys)} to write state variables, \texttt{pairit.event(name, data)} to record custom telemetry, and \texttt{pairit.done()} to signal completion. By writing to shared session state, custom HTML pages plug directly into matchmaking pools, autonomous agents, shared workspaces, routing conditions, and the central event-logging engine.

To protect platform integrity and participant privacy, custom components run inside a sandboxed iframe. This sandbox permits script execution but blocks direct network requests and prevents exposure of sensitive session tokens. Furthermore, researchers can declare \texttt{required: true} on the HTML component configuration. This option blocks the standard navigation button, forcing participants to complete the custom task before advancing in the study flow.

\section{Agent Execution, Tool Actions, and Collaborative Workspaces}

\paragraph{Agent Triggers and Dynamic Prompts.}
Pairit coordinates autonomous conversational partners and real-time editing environments directly through the declarative experiment configuration. Agents run on the server and connect to chat rooms through declared triggers. A trigger specifies when an agent activates: on every participant turn (\texttt{every\_message}), immediately upon entering the room (\texttt{on\_join}), or at regular message intervals (\texttt{\{every: N\}}). Researchers can also define conditional reply rules (\texttt{replyCondition}) so an agent evaluates whether an intervention is needed before responding. Agent system prompts support dynamic state interpolation (\texttt{\{\{session\_state.key\}\}}) as well as conditional prompt blocks (\texttt{prompts[].when}) that adjust instructions based on intake survey scores or assigned treatment conditions.

\paragraph{Tool Actions and Autonomous Interventions.}
Agents execute server-side tools defined with JSON Schema parameters to take actions beyond conversational chat. Built-in tools allow agents to update session state variables (\texttt{assign\_state}), conclude discussions when participants reach agreement (\texttt{end\_chat}), and edit shared documents in real time (\texttt{write\_workspace}). Through these tools, agents operate as active facilitators, bargaining counterparties, and co-authors rather than passive conversational responders.

\paragraph{Collaborative Workspace Modes and Scoping.}
The \texttt{live-workspace} component provides real-time co-authoring in two modes: freeform markdown editing and structured multi-field forms (such as text fields, numerical ratings, and textareas). Pages can use \texttt{layout: split} to display a workspace alongside chat in a two-column layout. The configuration scopes each workspace to an individual participant (\texttt{scope: participant}) or shares it across a matchmaking group (\texttt{scope: group}). Researchers set editing permissions (\texttt{editableBy: participant}, \texttt{agent}, or \texttt{both}) to test different divisions of labor and coordination dynamics between humans and AI agents.

\section{Comparison with Existing Platforms}

Online behavioral platforms differ in how they represent experimental logic and coordinate live participants. Survey platforms like Qualtrics support questionnaires and static branching, but cannot coordinate synchronous multi-party tasks, concurrent text editing, or dynamic AI interactions. Programmable behavioral frameworks such as oTree, Empirica, jsPsych, and nodeGame provide multi-player synchronization and browser-based cognitive tasks \citep{chen2016otree,almaatouq2021empirica,deleeuw2015jspsych,balietti2017nodegame}. Because these frameworks embed study logic in imperative Python or JavaScript code, integrating autonomous AI agents and shared workspaces requires custom application engineering. Deliberate Lab supports real-time human--AI group chat \citep{qian2025deliberatelab}, but does not integrate a lintable graph specification with concurrent shared workspaces and sandboxed custom HTML components. Pairit complements these systems by providing a declarative, graph-based architecture designed for multi-party human--AI collaboration (Table~\ref{tab:compare}).

\begin{table}[!ht]
\centering
\normalsize
\setlength{\tabcolsep}{6pt}
\renewcommand{\arraystretch}{1.55}
\begin{tabular*}{\textwidth}{@{\extracolsep{\fill}}>{\raggedright\arraybackslash}p{5.2cm}cccccc}
\toprule
Capability & Pairit & Empirica & oTree & jsPsych & nodeGame & \shortstack{Deliberate\\Lab} \\
\midrule
Declarative single-file specification & \checkmark & $\circ$ & --- & --- & --- & $\circ$ \\
Graph-based flow with conditional routing & \checkmark & $\circ$ & $\circ$ & $\circ$ & $\circ$ & $\circ$ \\
Built-in randomization (random / balanced / block) & \checkmark & \checkmark & \checkmark & $\circ$ & \checkmark & $\circ$ \\
Server-managed matchmaking with timeouts & \checkmark & \checkmark & $\circ$ & --- & \checkmark & \checkmark \\
Persistent multi-page chat groups & \checkmark & $\circ$ & --- & --- & $\circ$ & \checkmark \\
Native AI-agent integration & \checkmark & --- & --- & --- & --- & \checkmark \\
Collaborative shared workspace & \checkmark & --- & --- & --- & --- & $\circ$ \\
Sandboxed custom HTML components & \checkmark & $\circ$ & --- & $\circ$ & --- & $\circ$ \\
Lint / compile pipeline for experiment configs & \checkmark & --- & --- & --- & --- & --- \\
Structured event export (CSV / JSON / JSONL) & \checkmark & \checkmark & \checkmark & \checkmark & \checkmark & \checkmark \\
\bottomrule
\end{tabular*}
\caption{Capability comparison across selected online experiment platforms. \checkmark{} indicates first-class support, $\circ$ indicates partial or third-party support, and --- indicates no built-in support known to the authors at the time of writing. Entries reflect publicly documented features and are intended as a descriptive summary rather than a benchmark.}
\label{tab:compare}
\end{table}

\section{Event Logging and Custom Analytics}

Pairit automatically captures participant behavior and system lifecycle operations through a unified event-logging pipeline. When a participant interacts with a user interface element or the system shifts state, the platform emits a structured event payload. Every logged event contains a core context block consisting of the \texttt{sessionId}, \texttt{pageId}, and \texttt{componentId}. Beyond these standard system keys, researchers can record configurable custom metadata payloads by specifying \texttt{events.\{name\}.data} properties in the component configuration.

Table~\ref{tab:events} details the primary component events and their associated lifecycle triggers. To assist researchers in designing telemetry systems, the public library contains the \texttt{Component Events} example template. This template demonstrates how to structure custom event keys, monitor participant behaviors, and log real-time telemetry.

\begin{table}[!ht]
\centering
\normalsize
\setlength{\tabcolsep}{8pt}
\renewcommand{\arraystretch}{1.3}
\begin{tabular*}{\textwidth}{@{\extracolsep{\fill}}l l p{8.4cm}}
\toprule
Event Name & Component & Description and Context Payload \\
\midrule
\texttt{onClick} & \texttt{buttons} & Fires when a participant clicks a button; logs target navigation and custom click payload. \\
\texttt{onSubmit} & \texttt{survey} & Fires when a questionnaire is submitted; logs individual question answers and validations. \\
\texttt{onRequestStart} & \texttt{matchmaking} & Fires when a participant enters a matching pool; logs pool identifier and join timestamp. \\
\texttt{onMatchFound} & \texttt{matchmaking} & Fires when a group is successfully formed; logs matched group identifier and wait duration. \\
\texttt{onTimeout} & \texttt{matchmaking} & Fires when queue wait time expires; logs timeout duration and fallback routing. \\
\texttt{onCancel} & \texttt{matchmaking} & Fires if matchmaking is cancelled before a match or timeout. \\
\texttt{onMessageSend} & \texttt{chat} & Fires when a participant sends a chat turn; logs room identifier, text body, and sender metadata. \\
\texttt{onMessageReceive} & \texttt{chat} & Fires when a participant or agent receives a chat turn in the room. \\
\texttt{onStart} & \texttt{timer} & Fires when a page timer initializes; logs start times and total allocated durations. \\
\texttt{onWarning} & \texttt{timer} & Fires when a timer enters the warning threshold; logs duration remaining and styling shifts. \\
\texttt{onExpiry} & \texttt{timer} & Fires when a countdown expires; logs timeout events and subsequent navigation actions. \\
\texttt{onEdit} & \texttt{live-workspace} & Fires when a participant or agent modifies content; logs incremental keystroke-level diffs. \\
\texttt{onLoad} & \texttt{html} & Fires when a custom iframe finishes loading; logs browser readiness and client environment. \\
\texttt{onState} & \texttt{html} & Fires when a custom interface modifies state; logs written session state keys and values. \\
\texttt{onDone} & \texttt{html} & Fires when a custom task signals completion; logs finish events and termination context. \\
\bottomrule
\end{tabular*}
\caption{System-defined component events and lifecycle triggers. Each event automatically captures \texttt{sessionId}, \texttt{pageId}, and \texttt{componentId} alongside the listed metadata payloads.}
\label{tab:events}
\end{table}

\section{Manager CLI and Research Workflow}

The \texttt{pairit} command-line interface (CLI) coordinates the end-to-end experimental workflow from design to data retrieval. Researchers manage study deployments, configuration diagnostics, and data exports entirely through terminal commands. The standardized research workflow consists of five core phases (Table~\ref{tab:workflow}).

\begin{table}[!ht]
\centering
\small
\setlength{\tabcolsep}{6pt}
\renewcommand{\arraystretch}{1.25}
\begin{tabular*}{\textwidth}{@{\extracolsep{\fill}}c >{\raggedright\arraybackslash}p{5.1cm} p{8.2cm}}
\toprule
Phase & Command / Action & Description \\
\midrule
1 & Author YAML & Declare pages, components, matchmaking pools, and routing rules in one configuration file. \\
2 & \texttt{pairit config lint} & Validate the study schema locally before uploading. \\
3 & \texttt{pairit config upload {-}{-}config-id} & Compile the configuration and publish the study graph to the Manager Server. \\
4 & Share lab URL & Distribute \href{https://lab.pairium.ai/{configId}}{lab.pairium.ai/\{configId\}} to recruit participants into live sessions. \\
5 & \texttt{pairit data export} & Retrieve completed session logs and telemetry as flat analysis files. \\
\bottomrule
\end{tabular*}
\caption{End-to-end research workflow from configuration authoring through data export.}
\label{tab:workflow}
\end{table}

To prevent deployment errors, the CLI linter executes multiple structural checks. The linter validates the YAML configuration against a strict JSON schema, verifies that all page and component identifiers remain globally unique, and ensures that routing rules resolve to existing page targets. For studies utilizing custom HTML, the linter checks that all referenced HTML files exist locally, do not exceed the 1MB file size limit, and contain valid script structures.

Security and API access control also reside within this command-line workflow. When a study incorporates AI agents, the CLI accepts per-experiment API keys for external model providers. The platform encrypts these keys at rest on the database server. Because the platform does not provide fallback credentials, study configurations operate using only the researcher's encrypted keys. This architecture protects against unauthorized usage and ensures that researchers maintain direct, auditable control over their computational resources.

\section{Data Export and Protocol Auditing}

Pairit exports complete experimental records in standard CSV, JSON, and JSONL formats. The export process converts MongoDB collections into flat, structured files. Specifically, \texttt{pairit data export} produces six discrete relational files (Table~\ref{tab:export}).

\begin{table}[!ht]
\centering
\small
\setlength{\tabcolsep}{6pt}
\renewcommand{\arraystretch}{1.25}
\begin{tabular*}{\textwidth}{@{\extracolsep{\fill}}l p{10.8cm}}
\toprule
Export File & Contents \\
\midrule
\texttt{sessions} & Participant trajectory metadata, entry and exit timestamps, and final task statuses. \\
\texttt{events} & Fine-grained telemetry stream of all user interface interactions and system events. \\
\texttt{chat-messages} & Every conversational turn from chat rooms, including sender identifiers and exact timestamps. \\
\texttt{groups} & Matchmaking pool results, group structures, and collaborative room assignments. \\
\texttt{survey-responses} & Questionnaires, input values, and structured form responses. \\
\texttt{workspace-documents} & Keystroke-level edits, document contents, and revision histories. \\
\bottomrule
\end{tabular*}
\caption{Standard export files produced by \texttt{pairit data export} in CSV, JSON, or JSONL format.}
\label{tab:export}
\end{table}
These exported records preserve the exact temporal ordering of all actions, allowing researchers to reconstruct team deliberation, coordination, and task progress step by step. Experimental conditions with AI partners record complete model metadata alongside human behavioral data. Each agent transaction logs the model provider, model identifier, temperature and sampling hyperparameters, system prompt, conversation history, and raw token outputs. Explicit parameter logging keeps experimental protocols fully auditable across upstream model updates and behavioral drift. The single-file YAML configuration functions as an executable, auditable protocol. Before deployment, the Manager CLI lints the configuration against the platform schema, checking for missing page targets, dangling edges, invalid component properties, and malformed agent specifications. Researchers and reviewers can audit the entire experimental protocol by reading a single human-readable configuration file, without inspecting disparate server scripts or frontend application code.

\section{Interactive Template Library}

Pairit provides a comprehensive template library to accelerate study design and promote reproducible research. Scholars can inspect, customize, and run these configurations immediately within their web browser at \href{https://docs.pairium.ai/pairit/examples/}{docs.pairium.ai/pairit/examples}. The public documentation site runs these templates in a fully simulated frontend environment, allowing researchers to test complex multi-player flows and agent interactions without creating hosting accounts, configuring databases, or supplying API credentials.

The interactive library contains fourteen standard, runnable demonstration templates (Table~\ref{tab:templates}).

\begin{table}[!ht]
\centering
\small
\setlength{\tabcolsep}{6pt}
\renewcommand{\arraystretch}{1.2}
\begin{tabular*}{\textwidth}{@{\extracolsep{\fill}}l p{10.8cm}}
\toprule
Template & Description \\
\midrule
Hello World & Basic page nodes, text components, and simple navigation button transitions. \\
Mermaid & Dynamic flowchart visualizations using Markdown and Mermaid.js syntax. \\
Survey & Questionnaire construction, input validation, and automatic writes to session state. \\
Randomization & Simple, balanced, and block randomization on page entry. \\
AI Chat & Real-time chat room with a server-hosted AI partner on every participant message. \\
Multi-Chat & Shared or isolated chat history across pages via a common \texttt{groupId}. \\
Team Decision & Synchronous two-person matchmaking routed into a shared chat room. \\
AI Mediation & Autonomous facilitator that monitors discussion and intervenes when chat stalls. \\
Workspace & Collaborative co-authoring with real-time markdown and structured form fields. \\
HTML & Custom interface in a sandboxed iframe with bidirectional session-state API. \\
Timer & Visible countdowns, warning transitions, and automatic timeout routing. \\
Agent Triggers & Agent activation on room entry or at defined message intervals. \\
Conditional Agent & Dynamic agent prompts adapted from intake survey scores. \\
Component Events & Custom interaction telemetry from buttons, surveys, and workspaces. \\
\bottomrule
\end{tabular*}
\caption{Runnable demonstration templates available at \href{https://docs.pairium.ai/pairit/examples/}{docs.pairium.ai/pairit/examples}. Each template includes a live browser demo and downloadable configuration.}
\label{tab:templates}
\end{table}

\section{Ethics, Privacy, and Access Control}

Researchers configure ethical safeguards directly within the study specification. The configuration defines informed consent pages before interactive stages, controls whether partner identities and AI assistance are disclosed or blinded, and toggles data logging for chat and workspace activity. If synchronous matchmaking encounters participant dropouts or slow arrival rates, configurable timeout rules route stranded participants to fallback pages (such as solo tasks or exit surveys), ensuring full compensation without indefinite waiting.

API keys for commercial language model providers remain strictly on the Lab Server and never reach participant browsers. Browser clients receive only rendered responses, protecting model credentials and proprietary prompts from client-side inspection. We do not claim that self-hosting a multi-server real-time platform requires only a single command. Instead, we demonstrate working system readiness through the public documentation site, where researchers can run interactive templates and live demos in a browser without software installation or API credentials.

\section{Scale and Empirical Precedent}

An earlier deployment of the Pairit architecture powered a large-scale field experiment comparing human--human and human--AI teams \citep{ju2025collaborating}. That study recruited 2,234 Prolific participants who collaborated synchronously in pairs across shared workspaces and live chat to produce creative advertising campaigns evaluated in a live market test. The current platform preserves the same core functional primitives as that earlier deployment: real-time multi-party matchmaking, synchronous chat, concurrent workspace editing, server-managed AI agents, and automated experimental randomization.

High-throughput empirical trials confirm that these architectural primitives handle synchronized multi-party workflows at scale. Live participant loads validated real-time matchmaking, concurrent document editing, and server-managed AI interactions under heavy concurrent traffic. We do not claim that today's YAML configuration format runs earlier empirical studies without translation.

\section{Execution Risks and Open Questions}

The main risks to fully realizing this infrastructure are execution risks, not gaps in the declarative model itself.

\paragraph{External adoption and deployment.} The platform is operational and has supported peer-reviewed and forthcoming work, but wider use still depends on beta rollout to outside laboratories. Sharing a configuration link does not remove the need for institutional review, participant recruitment, model API credentials, and either managed hosting or a laboratory deployment. Whether executable protocols travel across research groups as routinely as survey instruments remains to be tested.

\paragraph{In silico pretesting.} We are developing simulated-participant pilots that run on the same declared configuration used for live sessions. That extension is not yet part of the released workflow. Even when available, language-model agents may reproduce some conversational patterns without capturing the full social dynamics of those live teams.

\paragraph{Upstream model dependence.} {\emergencystretch=3em Studies that embed commercial language models inherit provider access, pricing, and version changes. Pairit logs prompts, model identifiers, and outputs so protocols remain auditable, but replicating a published result may require the same model revision, which other laboratories may not be able to obtain.\par}

\section{Software Availability and Reproducibility}

The platform source code, CLI tools, and core server packages are public for inspection at \url{https://github.com/pairium/pairit}.

Documentation and interactive browser demos are at \url{https://docs.pairium.ai/pairit/}.

The stack uses Bun and Elysia on the backend, React with Vite and Tailwind CSS on the frontend, MongoDB for persistence, and server-sent events for real-time synchronization. Researchers run studies through the hosted platform.

We have built and fully operationalized the Pairit platform. During the initial alpha testing phase, we partnered with friendly research laboratories to deploy and refine the core system components. We are currently rolling out the beta release to external labs to support structural replications of collaborative studies. Insights from ongoing academic workshops continuously refine the public template library, addressing practical user friction and expanding experimental designs.

Participant-level data from cited empirical studies remain with their respective publications under original institutional review board protocols. The reproducible objects for this technical submission consist of public example configurations and standardized export schemas. The platform source is public for inspection.

\end{document}